\documentclass[proceedings,conferenceproceedings,accept,moreauthors,pdftex,10pt,a4paper]{Definitions/mdpi} 
\usepackage{amsmath}
\usepackage{braket}
\preto{\abstractkeywords}{\nolinenumbers}
\preto{\abstractkeywords}{\nolinenumbers}

\firstpage{1} 
\pubvolume{xx}
\issuenum{1}
\articlenumber{5}
\pubyear{2018}
\copyrightyear{2018}
\history{Received: date; Accepted: date; Published: date}

\Title{Insight into thermal modifications of quarkonia from a comparison of continuum-extrapolated lattice results to perturbative QCD$^{\dagger}$}

\Author{Anna-Lena Kruse $^{1}$* with H.-T. Ding$^{2}$, O. Kaczmarek$^{1,2}$, H. Ohno$^{3}$ and H. Sandmeyer$^{1}$}

\AuthorNames{Firstname Lastname, Firstname Lastname and Firstname Lastname}

\address{%
$^{1}$ \quad Fakult\"at f\"ur Physik, Universit\"at Bielefeld, 33615 Bielefeld, Germany\\
$^{2}$ \quad Key Laboratory of Quark and Lepton Physics (MOE) and Institute of Particle Physics, Central China Normal University, Wuhan 430079, China\\
$^{3}$ \quad Center for Computational Sciences, University of Tsukuba, Tsukuba, Ibaraki 305-8577, Japan }

\corres{Correspondence: alkruse@physik.uni-bielefeld.de}

\firstnote{Presented at Hot Quarks 2018 - Workshop for young scientists on the physics of ultrarelativistic nucleus-nucleus collisions, Texel, The Netherlands, September 7-14 2018}
\abstract{In this work, we strive to gain insight into thermal modifications of charmonium and bottomonium bound states as well as the heavy quark diffusion coefficient. The desired information is contained in the spectral function which can not be calculated on the lattice directly. Instead, the correlator given by an integration over the spectral function times an integration kernel is obtained. Extracting the spectral function is an ill-posed inversion problem and various different solutions have been proposed. We focus on a comparison to a spectral function obtained from combining perturbative and pNRQCD calculations. In order to get precise results, continuum extrapolated correlators originating from large and fine lattices are used. We first analyze the pseudoscalar channel since the absence of a transport peak simplifies the analysis. The knowledge gained from this is then used to extend the analysis to the vector channel, where information on heavy quark transport is encoded in the low frequency regime of the spectral function. The comparison shows a qualitatively good agreement between perturbative and lattice correlators. Quantitative differences can be explained by systematic uncertainties.}

\begin{document}


\section{Introduction}
\begin{figure}[t]
\centering
\includegraphics[width=0.49\textwidth]{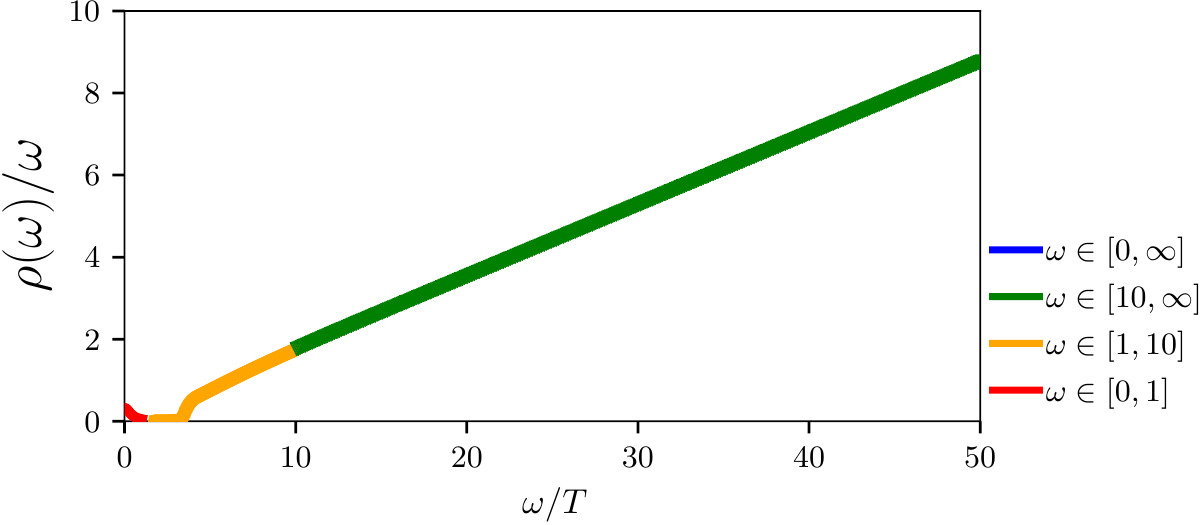}
\includegraphics[width=0.49\textwidth]{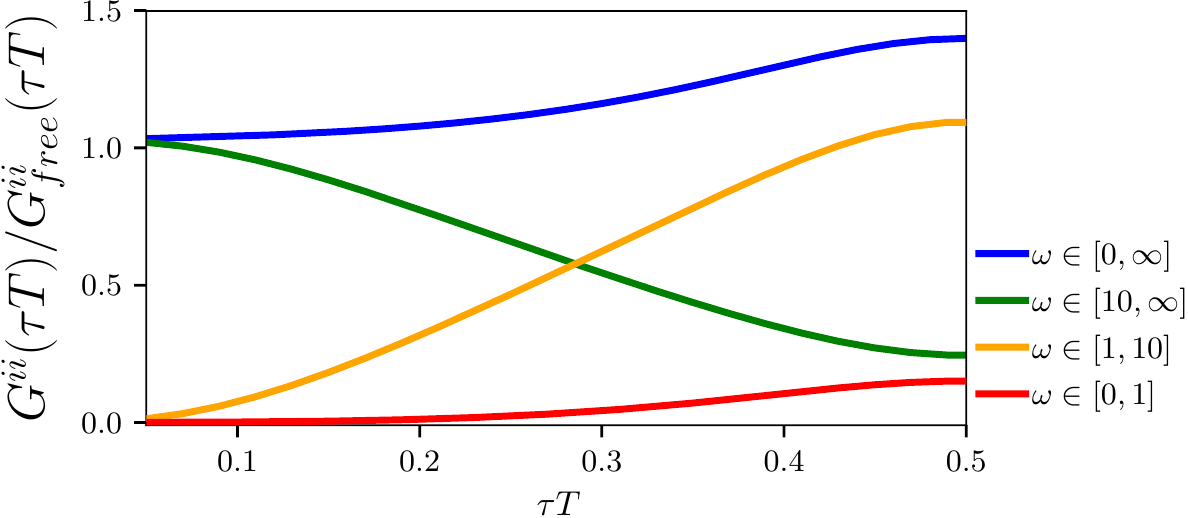}
\caption{The influence of different parts of the spectral function to the different regimes of the correlator.\label{fig-regionspfcorr}}
\end{figure}

Information on the in-medium properties like quarkonium bound states and heavy quark transport coefficients is contained in the spectral function. In perturbation theory, a spectral function can be constructed by combining different energy regimes. On the lattice, the spectral function can not be calculated directly. Instead, we compare lattice and perturbative correlators to gain insight into thermal modifications of quarkonia. Since the pseudoscalar channel does not contain a transport peak, it serves as an ideal probe for this comparison method. The knowledge obtained from pseudoscalar correlators can then be used to extend the analysis to the vector channel.\\
The connected (i.e. flavour non-singlet) imaginary time correlator in the pseudoscalar channel is given by
\begin{align}
G_{PS} (\tau)=M_{B}^2 \int_{\vec{x}} \braket{(\bar{\psi} i\gamma_5 \psi)(\tau,\vec{x})(\bar{\psi} i\gamma_5 \psi)(0,\vec{0})}_c, 
\end{align}
where $M_B$ is a bare quark mass and the vector correlator by
\begin{align}
G_{ii}(\tau)=\int_{\vec{x}}\braket{(\bar{\psi}\gamma_i \psi)(\tau,\vec{x}) (\bar{\psi}\gamma_i \psi)(0,\vec{0})}_c .
\end{align}
These correlators are related to the corresponding spectral functions through an integral equation,
\begin{align}
G_{PS,ii}(\tau)=\int \limits_{0}^{\infty} \frac{\mathrm{d}\omega}{\pi} \rho_{PS,ii}(\omega)K(\omega,\tau)\quad \text{with } \quad K(\omega,\tau)=\frac{\cosh(\omega(\tau-\frac{1}{2T}))}{\sinh(\frac{\omega}{2T})},
\end{align}
where the temperature $T$ is given by the inverse of the lattice extent. Due to the structure of the integration kernel $K(\omega,\tau)$, the low frequency regime of the spectral function influences the shape of the correlator at larger $\tau T$ while the high frequency part dominates the small distance part of the correlator. Fig.~\ref{fig-regionspfcorr} shows the contribution of different frequency regimes to the correlator. In the vector channel, a transport peak is expected at $\omega \approx 0$. Its contribution mainly influences the correlator at large $\tau T$, but has negligible effect on the low $\tau T$ regime. Still, the transport peak complexifies the analysis and it is thus easier to test the method in the pseudoscalar channel, where no transport contribution is present, and use the gained knowledge for the vector channel analysis.

\section{Continuum Extrapolation}
\begin{figure}[H]
\centering
\includegraphics[width=0.55\textwidth]{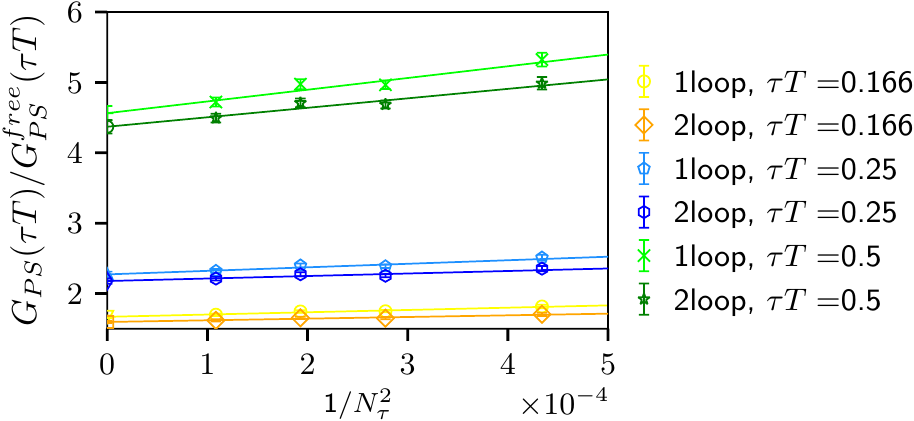}
\caption{Example of the continuum extrapolation in the pseudoscalar channel for three points in $\tau T$.\label{fig-contextr}}
\end{figure}

In order to realize the large lattices needed for this work, we employed the standard plaquette gauge action and used the quenched approximation with Clover-improved Wilson valence quarks. Five temperatures in the range from $0.75~T_c$ to $2.25~T_c$ were generated on four different lattice spacings and the correlators were continuum extrapolated according to the method described in \cite{ref-contextr}. The continuum extrapolation requires a renormalization of the correlators. For the pseudoscalar channel one- and two-loop perturbative renormalization constants are available. The continuum extrapolation is carried out using the two-loop expression and the difference is used to estimate a systematic error. In addition to the perturbative renormalization constants, non-perturbatively determined renormalization constants are available in the vector channel \cite{ref-renluescher}. The preferred option is to get rid of renormalization constants by building the renormalization independent ratio with the quark number susceptibility $\chi_q$. Since $\chi_q$ is given by the zeroth component of the vector correlator, it has the same renormalization that thus drops out in the ratio. We chose $\chi_q$ at the temperature $2.25T_c$ as normalization since the quark number susceptibility is more reliable at higher temperatures.\\
To ensure that the different lattices have the same vector meson mass, the correlators are mass-interpolated by fitting a quadratic exponential ansatz to the data points from different masses. The values of the fit function at the masses of $J/\psi$ and $\Upsilon$ at each point in $\tau T$ are taken as the new correlator. After the mass interpolation, the normalized correlators are continuum extrapolated with an ansatz quadratic in the lattice spacing. The continuum extrapolation is done for every point in $\tau T$ (see fig.~\ref{fig-contextr} for examples) and leads to the continuum results shown in section 4.
\section{The Perturbative Spectral Function}
The spectral function to which the lattice data should be compared is derived from an interpolation of different
calculations for different energy regimes as decribed in \cite{ref-ourpaper,ref-pertthreshold}. Well above the threshold, the ultraviolet asymptotics are known up to high loop order in both channels at zero temperature. For frequencies around the threshold, thermal contributions arise and the physics for non-relativistic quark masses can be described using pNRQCD with a perturbative real time static potential. The threshold behaviour and the vacuum asymptotics are interpolated in such a way that the transition as well as its first derivative is continuous. For frequencies way below the threshold, this description overestimates the spectral function and the spectral function is modified to contain the expected exponential suppression.  
\section{Comparison of Lattice and Perturbative Correlators}
\begin{figure}[H]
\centering
\includegraphics[width=0.32\textwidth]{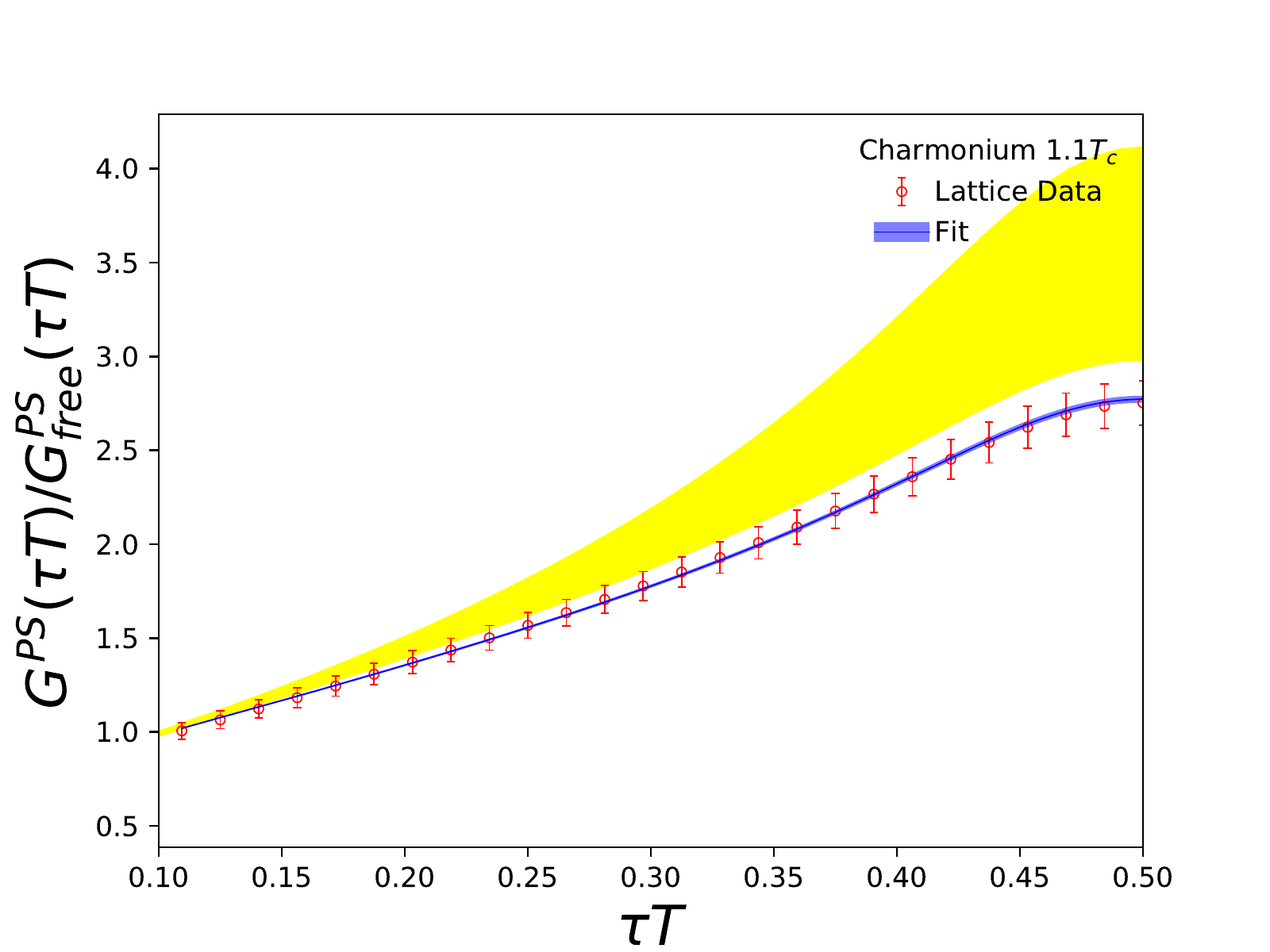}
\includegraphics[width=0.32\textwidth]{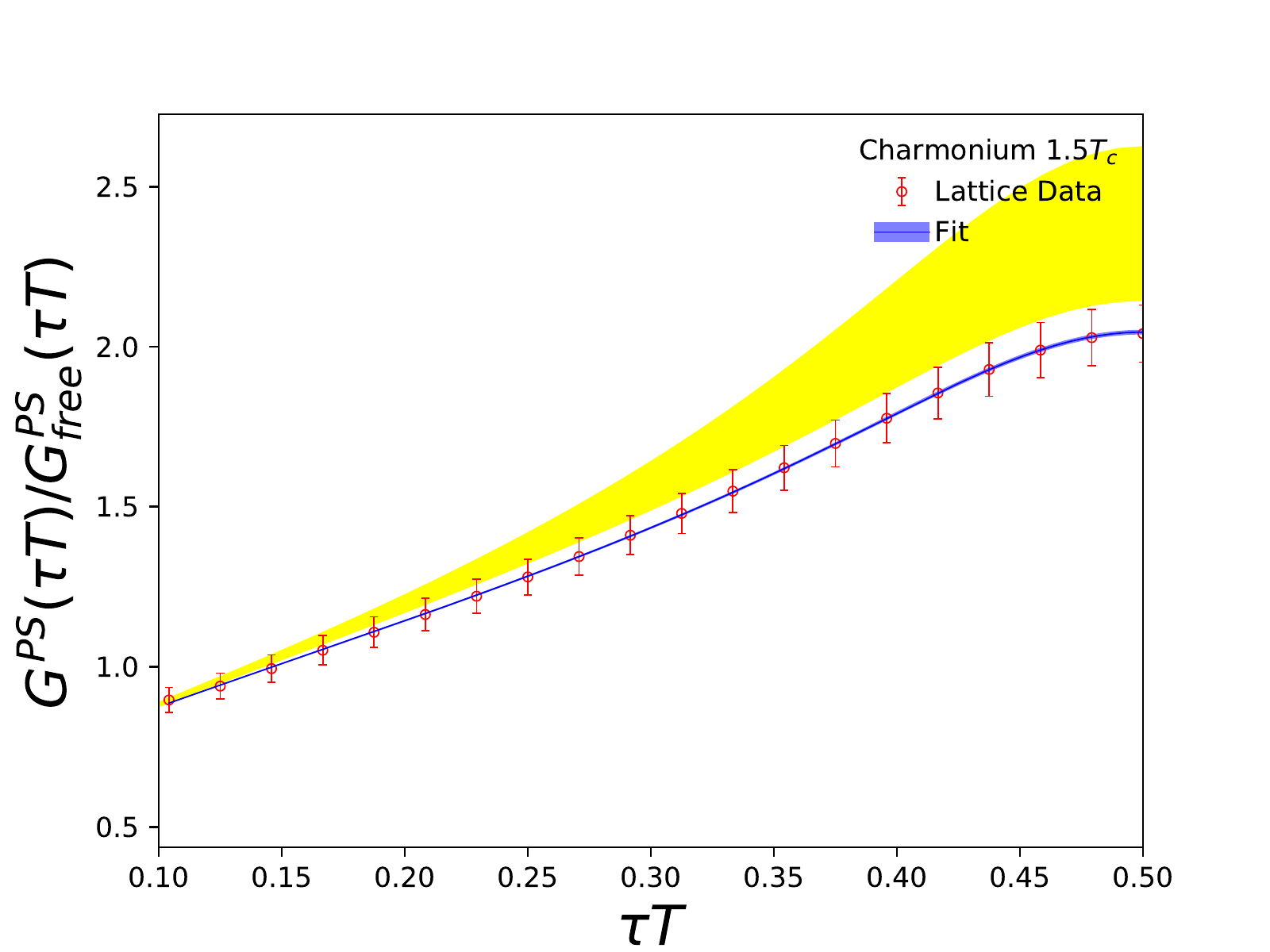}
\includegraphics[width=0.32\textwidth]{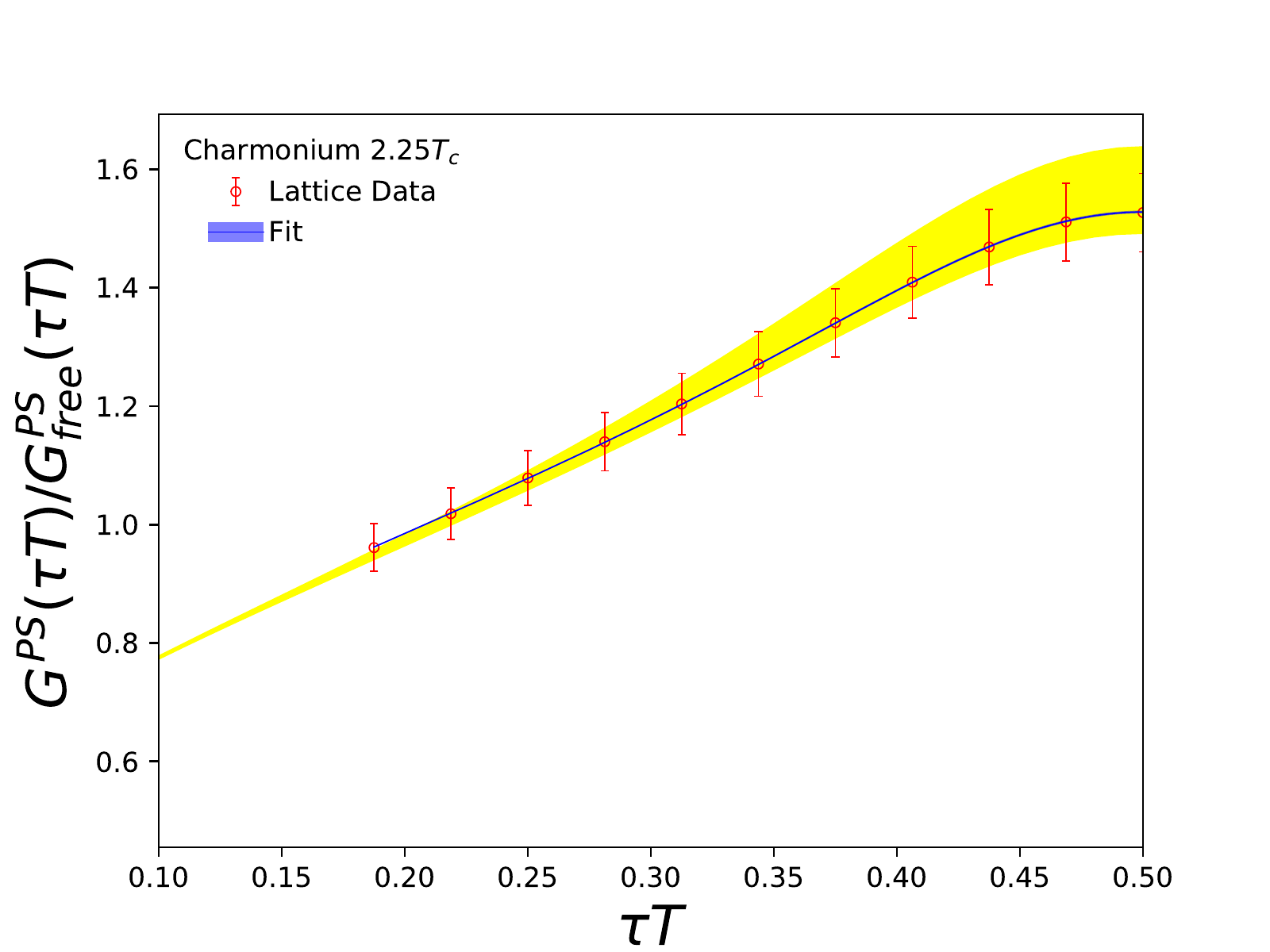}
\caption{Fit of the model spectral function (blue) to the lattice data in the charmonium pseudoscalar channel. The yellow line corresponds to the perturbative correlator calculated with a 10$\%$ uncertainty in the mass.\label{fig-PSpertfit}}
\end{figure}

A first look reveals a good qualitative agreement in the pseudoscalar channel. For a quantitative comparison, systematic uncertainties on both sides need to be taken into account. On the lattice side, the renormalization constants might be off. To account for this uncertainty, an overall factor $A$ is introduced. On the perturbative side, the relation between the pole mass and the $\overline{\text{MS}}$ mass is not exactly known which might result in a slightly smaller or larger threshold location. A mass shift $B$ is introduced, which can also account for any non-perturbative thermal mass shifts.\\
With these corrections, we construct a model spectral function $\rho^{model}(\omega)=A\rho^{pert}(\omega-B)$,
where $A$ and $B$ are variables in a fit to the lattice data. The fit results match the lattice data almost perfectly as shown in fig.~\ref{fig-PSpertfit}. In addition, $A$ is close to 1 and $B$ is small (see tab.~\ref{tab-PSpertfit}). Overall, a good agreement between lattice and perturbative correlators is observed in the pseudoscalar channel.\\
In the vector channel we apply the same method as in the pseudoscalar channel, i.e. building a model spectral function that is fitted to the lattice correlator. But here the transport peak complexifies the analysis. The shape of the transport peak can not be described perturbatively, only the constant contribution to the correlator can be estimated. Since the transport peak dominates the behaviour of the high $\tau T$ region of the correlator but has only little influence on the small $\tau T$ regime, we only fit the correlator for distances smaller than $\tau T=~$0.25. In the vector channel, the results at $2.25~T_c$ in fig.~4 also show good agreement between the lattice and perturbative correlators. The small deviation at large distances in the charmonium results may indicate a small contribution of a transport peak.
\begin{table}[h!]
\centering
\caption{Results from the fit of the model spectral function to the lattice data in the pseudoscalar channel from \cite{ref-ourpaper}.}
\label{tab-PSpertfit}
\tiny
\begin{tabular}{|c||c|c|c||c|c|c|}
\toprule
\tiny
 & \multicolumn{3}{c||}{\textbf{Charmonium}} & \multicolumn{3}{c|}{\textbf{Bottomonium}} \\\midrule
$T/T_c$ & $A$ &  $B/T$ & $\chi^2/d.o.f.$ & $A$ &  $B/T$ & $\chi^2/d.o.f.$\\ \midrule
1.1 & 1.04 &  0.52& 0.01 & 0.85 & -0.11 & 0.02\\
1.3 & 1.04 &  0.37& 0.01 & 0.87 & -0.13 & 0.04\\
1.5 & 1.02 &  0.33& 0.02 & 0.87 & -0.11 & 0.10\\
2.25& 1.06 &  0.16& 0.08 & 0.93 & -0.04 & 0.28\\
\bottomrule
\end{tabular}
\end{table}
\begin{figure}[H]
\centering
    \includegraphics[width=0.45\textwidth]{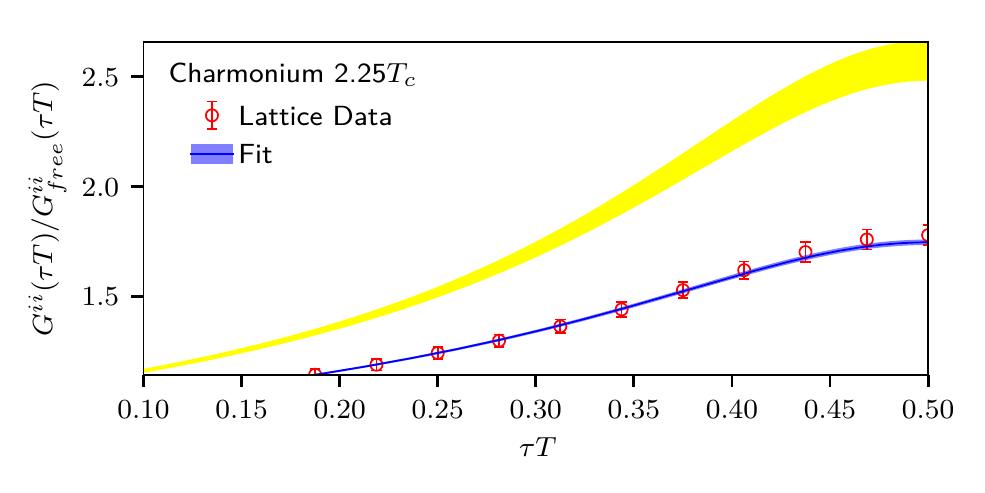}
    \includegraphics[width=0.45\textwidth]{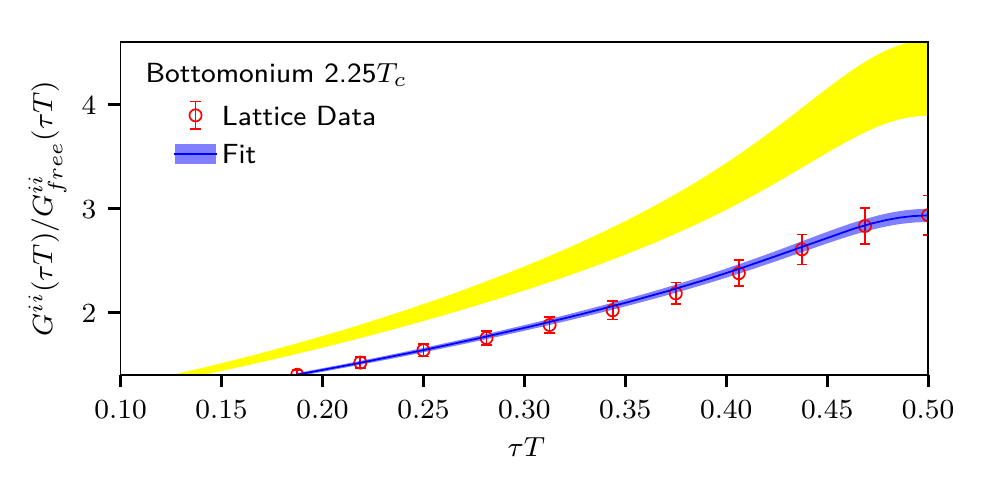}
\caption{Fit of the model spectral function (blue) to the lattice data up to $\tau T=0.25$ in the vector channel and the perturbative correlator (yellow).\label{fig-VVsmalltau}}
\end{figure}  
\section{Conclusions}
Overall, a good agreement between perturbation theory and lattice data is observed when taking systematic errors into account. It should be noted that this might partly be due to the smaller effective strong coupling in the quenched approximation. In the future, a full QCD study is required. The next steps are to crosscheck our results for the spectral function with the result of Bayesian reconstruction methods \cite{ref-sai2} and to fit suitable ans\"atze for the transport peak in the vector channel correlators.

\vspace{6pt} 




\funding{Funded by the Deutsche Forschungsgemeinschaft (DFG, German Research Foundation) – Project number 315477589 – TRR 211}

\reftitle{References}





\end{document}